\documentclass[jcp,aip,reprint,onecolumn]{revtex4-1}
\usepackage[colorlinks={true},dvipdfm]{hyperref}
\hypersetup{citecolor={blue}, filecolor={blue}, linkcolor={blue}, urlcolor={blue}}
\usepackage{anysize}
\usepackage{graphicx}
\usepackage{times}
\usepackage{color}
\marginsize{1 in}{1 in}{1 in}{1 in}
\usepackage{amsmath,amssymb}
\usepackage{dcolumn}% Align table columns on decimal point
\usepackage{bm}% bold math
\usepackage{enumerate}%\draft % marks overfull lines with a black rule on the right

\newcommand*{\ms}{\medspace}
\newcommand*{\brcl}{CH$_2$BrCl}

\newcommand*{\icl}{CH$_2$ICl}
\newcommand*{\ibr}{CH$_2$IBr}
\newcommand*{\cll}{CH$_2$Cl$_2$}
\newcommand*{\brr}{CH$_2$Br$_2$}
\newcommand*{\ii}{CH$_2$I$_2$}
\newcommand*{\brcll}{CHCl$_2$Br}
\newcommand*{\brrcl}{CHClBr$_2$}
\newcommand*{\clll}{CHCl$_3$}
\newcommand*{\brrr}{CHBr$_3$}
\usepackage{color}

\begin{document}
\title{Laboratory Transferability of Optimally Shaped Laser Pulses for Quantum Control}


\author{Katharine Moore Tibbetts}
\affiliation{Department of Chemistry, Princeton University, Princeton, NJ 08544, USA}
\affiliation{Current address: Department of Chemistry, Temple University, Philadelphia, PA 19122, USA}
\author{Xi Xing}
\affiliation{Department of Chemistry, Princeton University, Princeton, NJ 08544, USA}
\author{Herschel Rabitz}
\affiliation{Department of Chemistry, Princeton University, Princeton, NJ 08544, USA}


\date{\today}

\begin{abstract}
Optimal control experiments can readily identify effective shaped laser pulses, or ``photonic reagents'', that achieve a wide variety of objectives. For many practical applications, an important criterion is that a particular photonic reagent prescription still produce a good, if not optimal, target objective yield when transferred to a different system or laboratory, { even if the same shaped pulse profile cannot be reproduced exactly. As a specific example, we assess the potential for transferring optimal photonic reagents for the objective of optimizing a ratio of photoproduct ions from a family of halomethanes through three related experiments.} First, applying the same set of photonic reagents with systematically varying second- and third-order chirp on both laser systems generated similar shapes of the associated control landscape (i.e., relation between the objective yield and the variables describing the photonic reagents). Second, optimal photonic reagents obtained from the first laser system were found to still produce near optimal yields on the second laser system. Third, transferring a collection of photonic reagents optimized on the first laser system to the second laser system reproduced systematic trends in photoproduct yields upon interaction with the homologous chemical family. Despite inherent differences between the two systems, successful and robust transfer of photonic reagents is demonstrated in the above three circumstances. The ability to transfer photonic reagents from one laser system to another is analogous to well-established utilitarian operating procedures with traditional chemical reagents. The practical implications of the present results for experimental quantum control are discussed.
\end{abstract}

\maketitle
\section{Introduction}
Experimental optimal control of quantum systems is enjoying growing success since the advent of femtosecond pulse shaping technology, with widespread realizations for objectives as diverse as controlled molecular photodissociation \cite{Gerber02,Gerber2002,Weinacht2005b,Weinacht2006,hertel,melab2}, tailored high harmonic generation \cite{winterfeldt,pfeifer}, and controlled energy transfer and isomerization in biomolecules \cite{herek,Gerber05,Miller06,buckup}. Optimal control experiments (OCE) employ closed-loop adaptive feedback learning \cite{judson}, typically with a stochastic search algorithm (e.g., genetic algorithm) to identify a specially tailored shaped laser pulse, or ``photonic reagent'' \cite{hersch}, that induces the desired dynamical outcome in the target quantum system. In the last decade, more than one hundred successful OCE studies have been reported using closed-loop optimization \cite{constantin}.

The literature collectively indicates that optimal photonic reagents for many objectives can be readily identified \cite{constantin}. An important question is whether the prescription for a particular photonic reagent transforming a chosen system can similarly manipulate the same quantum system in a different laboratory with a like apparatus. In particular, the ability to transfer the control settings for an optimal photonic reagent to a different laser and pulse shaper while still achieving a good, if not optimal, outcome is of fundamental and practical importance. Transferability of specified chemical reagents and operating conditions from one laboratory to another is a cornerstone of chemistry, and the like behavior is naturally desired with laser control. Indeed, the same optimized radio frequency pulse sequences for nuclear magnetic resonance (NMR) applications are routinely employed by many laboratories to facilitate molecular structure elucidation \cite{spiess}. Effective transfer of photonic reagents would establish advantageous analogous behavior between photonic and chemical reagents, as well as facilitate the implementation of photonic reagent control for practical applications. As an illustration, this work explores the extent to which photonic reagents can effectively control the fragmentation of a family of halomethane compounds when implemented on a different laser system and pulse shaper. The example provides a good test of transferability, as the physical process is quite nonlinear with respect to the photonic reagent, possibly making the transfer sensitive to even small differences in the two laser systems. 

Typically, generating an arbitrarily shaped photonic reagent requires a femtosecond laser system and a pulse shaper, e.g., liquid crystal modulator (LCM) or acousto-optic modulator (AOM). The full prescription of a photonic reagent includes the laser pulse energy along with the spectral bandwidth, phase, and amplitude. For a given laser system with fixed energy and spectral bandwidth with only phase shaping employed, the spectral phase completely prescribes the photonic reagent and may conveniently be transferred to a second system. In this case, photonic reagent transfer requires specifying or calibrating the control settings on the second pulse shaper to reproduce the original spectral phase. { Characterization of the corresponding output shaped pulses, for example via FROG \cite{trebino}, SPIDER \cite{iaconis}, or SEA TADPOLE \cite{coughlan}, could verify production of the same output shaped pulse. However, coupling of the spatial and temporal profiles of the laser pulse arising from distortion introduced by the optical elements of the shaper can result in distinct spatial pulse profiles, even for nominally the same temporally shaped pulse \cite{pretzler,wefers, tanabe}. Such subtle differences are magnified when the pulse is focused after traveling a significant distance from the shaper \cite{stolow, coughlan}, as different spectral components can focus at distinct spatial locations within the beam waist. These limitations make it extremely difficult to deliver identical spatio-temporal pulse profiles to the target system, even when nominally the same photonic reagent is prescribed. 

Our experiments take the more practical perspective of assessing photonic reagent transferability directly for the target application by measuring the response of a chemical system to photonic reagents prepared on two different laser systems. As the same model of LCM and similar shaper optics are employed, photonic reagent transfer is performed simply by transferring the phase mask on the LCM pixels from one system to another. Assessing photonic reagent transfer success purely via the target control objective is particularly relevant to chemistry applications where the primary goal is to produce the desired product under any reasonable laboratory conditions. Successful control with transferred photonic reagents in this context would indicate a desirable degree of robustness in the control process to differences in the output laser pulse profiles, as no special effort is made to produce the same characterized pulse shapes on the two laser systems.

This work considers selective fragmentation of halomethanes detected by time-of-flight mass spectrometry (TOF-MS) as a test of photonic reagent transfer for controlling chemical reactions.} The halomethanes form a homologous chemical family that has been shown to exhibit systematic trends in ionized photoproduct yields upon interaction with families of shaped laser pulses \cite{melab1,melab2}. The latter works considered the ionized photoproducts resulting from cleavage of a carbon-halogen bond, e.g., methyl halide fragment ions CH$_2$X$^+$ or CH$_2$Y$^+$ and halogen ions X$^+$ or Y$^+$ resulting from the fragmentation of CH$_2$XY, where X and Y denote halogen atoms. The relative yields of these photoproducts were found to vary systematically with both halogen composition of the parent molecules and the laser pulse shape \cite{melab1,melab2}. The compounds \cll, \clll, \brcll, \brrcl, \brrr, \brr, \brcl, \icl, \ibr, and \ii\ms were considered in the latter studies and will also be employed for this work. For CH$_2$XY, and similar analogs in the halomethane family, we will focus on enhancing the fragment ion X$^+$ over that of CH$_2$X$^+$, where C$-$X is the stronger carbon-halogen bond. For this purpose, the optimal photonic reagent prescriptions (which will be referred to as ``photonic reagents'' in the remainder of the paper) identified in Ref. \cite{melab2} on a particular laser system (Coherent Legend, called ``System I'' in the remainder of this paper) will be implemented on a different laser system (KM Labs Dragon, called ``System II'') in order to assess their efficacy when transferred. The opportunity arose to test photonic reagent transferability upon the necessity of moving the TOF-MS setup to System II from System I when relocating our laboratory. As this work aims to explore the feasibility of photonic reagent transferability, additional tests were performed by conducting new optimization experiments on System II and comparing their outcome with the I$\to$II transferred photonic reagents. However, as the TOF-MS apparatus was moved to System II after the completion of the experiments in Ref. \cite{melab2}, it was not possible to further test the inverse transfer II$\to$I.

The remainder of this paper is structured as follows. Section \ref{expt} discusses the experimental conditions that are expected to impact the effective transfer of photonic reagents for typical experimental applications, with emphasis on the issues that arise for the TOF-MS experiments performed here. Section \ref{result} presents the results of three sets of experiments showing that transfer of photonic reagents from one laser system to another produces qualitatively similar experimental yields. The study also revealed some key laser system variables for attaining successful transfer of photonic reagents. Finally, Section \ref{disc} presents a discussion and concluding remarks.

\section{Experimental Methods}\label{expt}
This section considers the experimental conditions that may influence the efficacy of optimal photonic reagents upon transfer from one laser system to another with a particular focus on the relevant parameters used in our experiments. Beginning with laser pulses of similar energy, bandwidth, and stability from the output of the two laser systems is a necessary, but not sufficient, condition for successful transfer of photonic reagent LCM masks because the spatio-temporal profiles of the respective shaped laser pulses may still be different. Further issues arise with laser pulse delivery, particularly for the present TOF-MS experiments, as the laser pulses must be tightly focused to control the target species. Differences in the detection apparatus between the two laser systems may also play a role in some experiments.

\subsection{Generation of ultrafast laser pulses}
To expect any degree of transferability of prescribed photonic reagents from one laser system to another, the unshaped laser pulses generated by the two laser systems should have similar energy, spectral bandwidth, etc. A further important criterion is sufficient stability of the laser systems, as energy and phase noise in the output unshaped laser pulses can have a significant impact on the outcome of optimal control experiments \cite{roth1,roth2}, especially when the objective depends on quantum coherence processes driven very non-linearly by the photonic reagents. Other experimental issues, such as temperature and humidity control in the laboratory, as well as the stability of the optical tables, can play a role in determining reproducible laser performance. These issues can also affect analogous experiments with chemical reagents in different laboratories.

The operating characteristics of the two amplified Ti:Sapphire laser Systems I and II used in our experiments are presented in Table \ref{param1}, which includes the ranges of pulse duration and bandwidth observed from typical operating conditions of each system. The transfer experiments were performed with no further attempt to refine these variables. The two laser systems have comparable transform limited pulse durations, as measured by second harmonic FROG, although the bandwidth of System I is slightly higher than that of System II. The most significant difference between the two systems is the blue-shifted spectrum of System II compared to System I, as evident by the center spectral wavelengths reported in Table \ref{param1}. The experiments will explore whether the control yields achieved on the two laser systems are affected by shifting LCM mask generated on System I such that the phase function placed on the LCM of System II lies at the equivalent spectral range (i.e., shifted to the red portion of the spectrum on the LCM of System II). Although the pulse energy and spectral bandwidth ranges of System II lie slightly below System I, the typical ranges of operating parameters suggest that subsequent pulse shaping on System II may produce similar output shaped pulses.

\subsection{Shaping the laser pulses}
A primary challenge for transferring photonic reagents from one laser system to another is to generate similar pulse profiles from the same photonic reagent prescription, assuming that the unshaped laser pulses from Systems I and II are reasonably similar. Even having the same pulse shaper characteristics may not produce the same output shaped laser pulse structures due to subtle variations in the pulse shaper alignment leading to unmanaged spatio-temporal coupling of the laser pulse arising from shaping \cite{pretzler,wefers, tanabe}.
 
The pulse shapers on Systems I and II had their own optical components, although both employed the same model of LCM (CRI, SLM-640). The values of salient pulse shaper parameters for the two systems are compared in Table \ref{param2}. The small difference in the reported wavelength (nm) per LCM pixel is due to the distinct resolution of the respective diffraction gratings in the pulse shapers (1600 lines/mm on System I and 1400 lines/mm on System II). Overall, the similarity between the pulse shaper parameters for Systems I and II suggest that similar, but not identical, output laser pulse shapes could be produced when the same phase mask is applied. Although a more refined transfer of photonic reagents could have been performed by accounting for slight differences in spectral calibration of the two lasers, the simple transfer of a phase mask performed adequately in these experiments. 

\subsection{Shaped laser pulse delivery}\label{deliver}
In addition to the parameters of the respective laser systems and pulse shapers, the optical setup used to deliver the shaped laser pulses may cause discrepancies between the shaped pulses delivered to the target quantum system. The details are unique for each type of experiment regarding which parameters for shaped pulse delivery contribute the most to achieving effective transferability of photonic reagents. Here, we report the experimental parameters related to focusing of the shaped laser pulses into the TOF mass spectrometer in our experiments.

Spatio-temporal coupling in the beam focus can result in varying temporal pulse profiles at different locations over the cross-section within the Rayleigh length, particularly for pulses with complex phase shapes \cite{coughlan, stolow}. Residual spatial chirp in the unfocused laser beam can also cause different spectral components to focus at slightly different spatial locations within the beam waist. Even for identical shaped pulses, the path length from the pulse shaper to the focal lens can influence the temporal and spatial profile of the resulting focused beam \cite{stolow}. These effects can be normalized out to some extent by employing the same focusing conditions for all experiments conducted on one laser system, but with two distinct laser systems, slight deviations in output laser pulse shapes can result in significantly different beam profiles at the respective foci.

The experimental parameters affecting the delivery of focused laser pulses are provided in Table \ref{param3} for Systems I and II. The path lengths of the beam from the respective pulse shapers to the focusing lens differed by $\sim 15$cm on the respective systems out of lengths greater than 300cm. Although this difference is small, subtle discrepancies in the collimation quality of the output beam from each amplifier, as well as in the pulse shaper alignments between Systems I and II may result in differences in the divergence of the output beams from the pulse shapers. Thus, depending on the collimation quality of the respective beams out of the pulse shapers, the small path length difference possibly could have an effect on the respective focused beams. A plano-convex focusing lens of $f=20$cm was employed for experiments on both Systems I and II. The focal spot sizes were measured with a knife edge and found to be 40$\mu$m and 45$\mu$m for the Systems I and II, respectively, producing a $\sim20\%$ higher intensity on System I. Intensity variation is of primary importance in TOF-MS experiments \cite{gibson,xiazhang,liu2003,wang2006,hankin}. The photonic reagent transfer experiments performed in Section \ref{result} will reflect the cumulative {\it discrepancies} in the estimates of Tables \ref{param1}, \ref{param2}, and \ref{param3}, which are { solely based on easily assessed observables in the target chemical system.}

\subsection{Detection apparatus}
Although the detection apparatus does not influence the shaped laser pulse, it can play a role in assessing successful photonic reagent transfer, where each type of detector is expected to present its own issues that must be addressed. For example, successful implementation of optical detection on two different experimental setups must take into account the beam alignment, photon detector sensitivity, spectral filtering issues, etc. Here, we discuss the detection features inherent for TOF-MS detection of ions formed by interaction of a gas-phase molecular sample with a focused shaped laser pulse. In our experiments, the same TOF-MS chamber (Jordan TOF) was employed for detection on both laser systems, as it was moved from System I to System II for the practical reasons explained in the Introduction. Although this situation to some extent mitigates issues changing the entire experimental apparatus in order to assess photonic reagent transferability, it is unlikely that a properly set up TOF-MS detector would become the main source of any differences observed in the experimental results as variations in the laser, pulse shaper, and focusing parameters are likely to play a larger role.

\begin{figure}[htbp]
\begin{center}
\includegraphics[width=8.5cm]{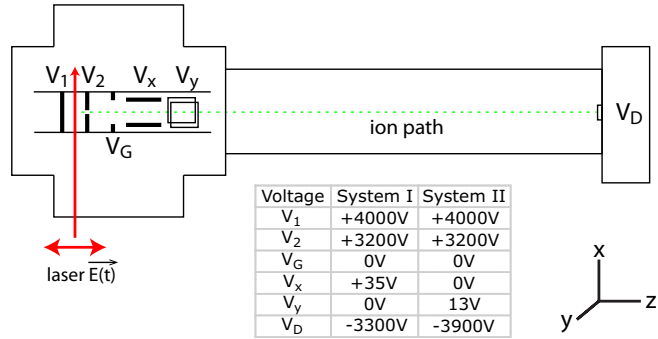}
\end{center}
\caption{\footnotesize Schematic diagram of the TOF apparatus with a table showing the values of the voltages placed on each plate indicated in the diagram. The laser path is along the x axis and its polarization is along the z axis, which is the ion flight path direction. \label{tof}}
\end{figure}

A schematic diagram of the apparatus, along with a table showing the voltages placed on each labeled plate, is presented in Figure \ref{tof}. The laser focus was centered between the repeller plate and extraction grid at respective voltages V$_1$ and V$_2$, which were separated by 1cm. The ions then pass through an acceleration grid at voltage V$_\text{G}$, followed by two sets of plates at voltages V$_\text{x}$ and V$_\text{y}$ that impart fields along the x and y axes to correct for any imperfections in the electric field defined by V$_1$, V$_2$, and V$_\text{G}$. The voltages on these plates were optimized to produce the greatest overall ion signal and were similar for both laser systems. Finally, the ions pass through a 1-meter field-free flight tube to a dual 18mm microchannel-plate detector (Jordan, C-701) with applied voltage V$_\text{D}$. The higher voltage on System II reported in Figure \ref{tof} was used to enhance the overall ion signal, as it was lower than on System I for the same detector bias because a pre-amplifier (Stanford, SR445A) was employed on System I; the preamplifier was subsequently found to distort the baseline of the TOF spectra in System I and was not employed for the experiments on System II. The resulting ion signal was measured with a digital oscilloscope (LeCroy, 104MXi) for both laser systems.

The geometry of the ion extraction optics determines both the collection efficiency of the ions and the spatial regions of the focused laser beam from which ions are collected. Failure to constrain the spatial volume for ion collection has been shown to result in broad spatial averaging of ions collected from regions of the laser beam with significantly different laser intensity, which complicates interpretation of mass spectral data \cite{hansch,jones,hankin,strohaber}. For the present experiments on both laser systems, a pinhole of 0.5mm in diameter in the V$_2$ plate ensured collection of ions only from the laser focal region, as shown schematically by the small hole on the V$_2$ plate in Figure \ref{tof}. With our beam focusing parameters, the Rayleigh length of the focused beam is $\sim5$mm, so the pinhole ensures collection only from within the region of the highest laser intensity along the x and y axes. This pinhole setup is widely used in order to minimize spatial averaging effects \cite{hansch,jones,hankin,strohaber}. The use of the same pinhole for experiments both on Systems I and II is expected to normalize out most spatial averaging differences. The molecular samples \cll, \clll, \brr, \brrr, \brcl, \icl, \ibr, \ii, \brcll, and \brrcl\ms used for the experiments on Systems I and II were purchased from Sigma-Aldrich and employed without further purification. Each sample was introduced to the TOF-MS chamber through the same effusive leak valve (MDC, ULV-150). The base pressure of the vacuum chamber was 10$^{-8}$ torr, and upon adding the sample, the pressure was typically maintained at 10$^{-6}$ torr for the experiments. 

\subsection{Referencing the LCM pulse shaper for transfer}\label{tlnorm}
An optimal ``photonic reagent'' is represented by the LCM mask identified by a search algorithm to control a particular objective. In order to have any reasonable expectation of reproducing a photonic reagent on the same laser system, let alone transfer the photonic reagent to a different laser system, the LCM mask specifying the photonic reagent must be referenced to the phase profile for a known physical process. This reference is commonly chosen as second harmonic generation (SHG) or non-resonant two-photon absorption (TPA), which are known to optimize with the transform-limited (TL) laser pulse having a flat spectral phase \cite{baumert,brixner}. While second-order dispersion in the amplifier output can be removed by a dual-grating compressor, higher-order phase distortions can remain. In order to remove this residual dispersion, the laser output TPA measured by a two-photon diode (Thorlabs) is optimized using the search algorithm (c.f., Section \ref{alg}) prior to optimizing the target objective. This TL reference phase is then added to subsequently identified LCM masks to normalize  them with respect to residual amplifier dispersion \cite{wolpert,roslund,melab2,dantusmiips}. In our experiments, the TL reference phase was typically updated daily.

\subsection{Search algorithm and control objective in the present experiments}\label{alg}
The main goal of this work is to explore photonic reagent transferability. As a cross-check of the photonic reagent transfer process, optimal photonic reagents discovered using System I in Ref. \cite{melab2} are compared with new optimal photonic reagents identified on System II by comparing the recorded TOF spectra on System II. A genetic algorithm (GA) \cite{goldberg} with the same parameters as in Ref. \cite{melab2} is employed in this work to identify new optimal photonic reagents on System II. Phase-only pulse shaping is employed in all of the experiments, with the LCM pixels specified in a polynomial spectral basis
\begin{equation}
\Phi(\omega)=A(\omega-\omega_0)^2+B(\omega-\omega_0)^3+C(\omega-\omega_0)^4,\label{poly}
\end{equation}
where the parameters $A$, $B$, $C$, and $\omega_0$ are subject to optimization by the GA. 

The control objective from Ref. \cite{melab2} of optimizing the production of the halogen ion photoproduct from carbon-halogen bond cleavage is considered here. The cleavage of a carbon-halogen bond in halomethane compounds results in a halogen atom and an associated methyl halide fragment, either or both of which may be ionized. Under interaction with TL pulses, the methyl halide ion product is favored due to its lower ionization potential \cite{wang2006,melab1,melab2}; the present control objective is to enhance the formation of the minor halogen ion product, which was found to be amenable to optimal control in Ref. \cite{melab2}. In the experiments of Ref. \cite{melab2}, the control objective was quantified as the ratio of the integrated signals of the desired halogen ion $S_1$ to the undesired methyl halide ion $S_2$, as indicated for each of the ten target compounds in Table \ref{fit2}. The integrated ion signals $S_1$ and $S_2$ over the full width of the respective peaks in the TOF spectrum defined the yield $J$
\begin{equation}
J=\frac{S_{1}}{S_2}.\label{obj2}
\end{equation}
 In order to prevent a singularity in $J$ at small values of $S_2$, any shaped pulse producing a value of $S_2$ below a specified threshold was given a objective yield of $J=0$ in the GA. Due to differences in intensity at the laser focus between the two systems (c.f., Section \ref{int}), the $S_2$ threshold utilized on System I was found to cause difficulties in transferring some optimal pulses to System II, as discussed in Section \ref{homo}.

In summary, the two Systems I and II are similar in a fashion envisioned to be the case in many laboratories. The observed system parameters appear to be sufficiently alike to expect some degree of similarity in the delivered photonic reagents, unknown spatio-temporal characteristics notwithstanding. Thus, the present tests are expected to be a reasonable assessment of the prospect for performing routine photonic reagent transferability { based only on the observed responses of the target chemical system.}

\section{Results}\label{result}
The results of three sets of experiments are presented demonstrating that the phase masks prescribing photonic reagents identified on System I can produce similar experimental results when implemented on System II. Following a qualitative calibration of the focal laser intensity on the respective laser systems in Section \ref{int}, Section \ref{param} investigates the response of the objective ratio in Eq. (\ref{obj2}) for \brcl\ms to systematic variation of the second and third-order spectral phase implemented on Systems I and II. Section \ref{opt} compares the efficacy of optimal photonic reagents identified on Systems I and II for \brcl\ms and \icl. Finally, Section \ref{homo} shows that systematic trends across the entire molecular family that were observed on System I in Ref. \cite{melab2} are qualitatively reproducible on System II.

\subsection{Comparison of relative laser intensity}\label{int}
In order to interpret the subsequent assessments of transferring photonic reagents from System I to System II, the relative focal laser intensities upon operation with each system must be determined. While sophisticated optical detectors may be employed for this purpose \cite{coughlan}, comparing the features of the TOF spectra obtained on each laser system under excitation with the TL pulse provides a qualitative calibration of the relative laser intensities \cite{wang2006,hankin,liu2003,xiazhang,gibson}. Although estimated peak laser intensities were calculated using data on the pulse energy, duration, and beam geometry in Section \ref{deliver}, it is important to test these estimates using a measurement of laser intensity that is directly based on the TOF-MS experiments performed in this work to assess photonic reagent transferability. We will show that the {\it actual} laser intensity, as measured by the TOF spectra, is significantly lower than the estimated intensity on System II as compared to System I.

\begin{figure}[htbp]
\begin{center}
\includegraphics[width=8.5cm]{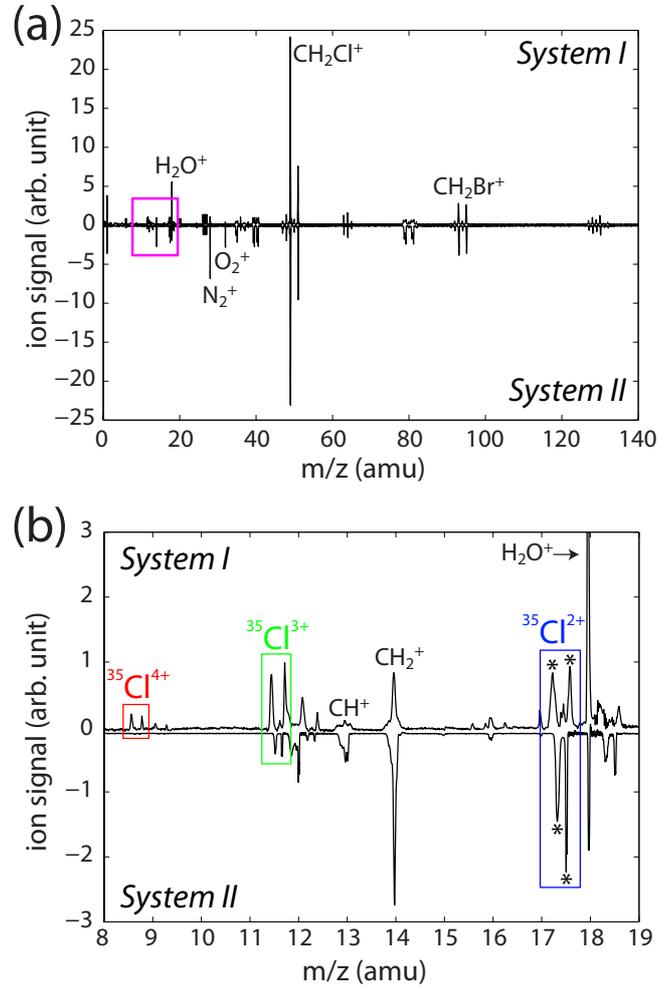}
\end{center}
\caption{\footnotesize TOF spectra from \brcl\ms recorded on System I (top of each plot) and System II (bottom of each plot) with their respective TL pulses. (a) Full TOF spectra. (b) Magnification of the region showing Cl$^{2+}$ through Cl$^{4+}$ (in the magenta box on (a)). The multiply-charged chlorine ions are shown by the blue, green, and red boxes for $^{35}$Cl$^{2+}$, $^{35}$Cl$^{3+}$, and $^{35}$Cl$^{4+}$, respectively. While significant Cl$^{4+}$ is observed in the spectrum from System I, this ion is not observed on System II. The distance between the two peaks marked with an asterisk on the Cl$^{2+}$ signal in (b) denotes the kinetic energy release from Coulomb explosion and is clearly greater on System I. \label{tl}}
\end{figure}

For halomethanes, two spectral features of halogen ions are highly dependent on the intensity and may be used for determining the relative intensities produced on Systems I and II in our experiments. First, higher charge states of halogen ions are observed for a higher laser intensity \cite{liu2003,xiazhang,wang2006}. Second, both singly- and multiply-charged halogen ions are formed, in part, by Coulomb explosion of a multiply-charged precursor ion. The corresponding halogen ion signals have at least two peaks corresponding to the ion's initial kinetic energy oriented toward (left peak) or away from (right peak) the detector (c.f., Figures \ref{tl} and \ref{tofoldnew}) upon Coulomb explosion \cite{gibson,xiazhang,liu2003,melab1,wang2006}. The kinetic energy released from Coulomb explosion is proportional to the distance between the two ion peaks, and is known to increase with laser intensity \cite{gibson,liu2003}. Thus, comparing both the highest visible charge state of each halogen and the distance between the two Coulomb explosion peaks in the TL spectra can determine the relative intensities produced on the respective systems. 

The relative intensities produced on Systems I and II were determined by inspection of the TOF spectra of \brcl\ms upon interaction with the TL pulse. These TOF spectra are shown in Figure \ref{tl}(a) (full spectra) and (b) (magnification of the region containing multiply-charged chlorine ions), with the spectra from System I plotted on the upper half of each plot, and the (inverted) spectra from System II plotted on the lower half of each plot. While the spectra in Figure \ref{tl}(a) look qualitatively similar, subtle differences become apparent upon magnification in Figure \ref{tl}(b), where the ion signals from Cl$^{2+}$, Cl$^{3+}$, and Cl$^{4+}$ are indicated by the blue, green, and red boxes, respectively. First, the distance between the two Coulomb explosion peaks in the Cl$^{2+}$ and Cl$^{3+}$ ions is much greater for System I, as indicated by the asterisk ($*$) labeling the two peaks from Cl$^{2+}$ in the blue box. Second, there is a significant yield of Cl$^{4+}$ visible in the spectrum taken on System I (top), but a negligible yield of this species is observed in the spectrum taken on System II (bottom). Cl$^{4+}$ was observed in \brcl\ms spectra taken on System I with reduced-energy TL pulses producing as little as $\sim50\%$ of the laser intensity reported in Table \ref{param1} (not shown). These results suggest that the maximum intensity produced on System II is no more than $\sim50\%$ of the intensity produced on System I, which is significantly less than the $\sim 80\%$ of the System I intensity expected based on the experimental parameters in Tables \ref{param1} and \ref{param2}. 

The discrepancy in actual focal intensity between the two laser systems may have contributions from small differences in the spatial chirp of the output beam from the respective pulse shapers \cite{coughlan,stolow}. The experimental results presented below conducted on System II will be discussed within the context of the difference in the laser intensities observed on Systems I and II. 

\subsection{Response of \brcl\ms to systematic photonic reagent variation}\label{param}
To assess the response of \brcl\ms to systematic variation of the laser pulse structure, we implement the same set of spectral phases on Systems I and II by scanning the two parameters $A$ and $B$ in Eq. (\ref{poly}), with $C=0$ fixed. The resolution limits of the LCM permitted variation of the linear chirp coefficient $A$ over the range of approximately $\pm2\times10^4$ fs$^2$, and of the cubic phase coefficient $B$ over the range of approximately $\pm4\times10^5$ fs$^3$. The slight difference in the wavelength/pixel resolution between the System I and System II shapers (c.f., Table \ref{param1}) does not significantly change the resolution limits. The center of the polynomial $\omega_0$ in Eq. (\ref{poly}) was taken to be the center pixel for both shapers, thus not correcting for the 9nm difference in central spectral frequency from the two laser sources. The effects of correcting for the difference in center wavelength will be examined in Section \ref{opt}.

\begin{figure}[htbp]
\begin{center}
\includegraphics[width=8.5cm]{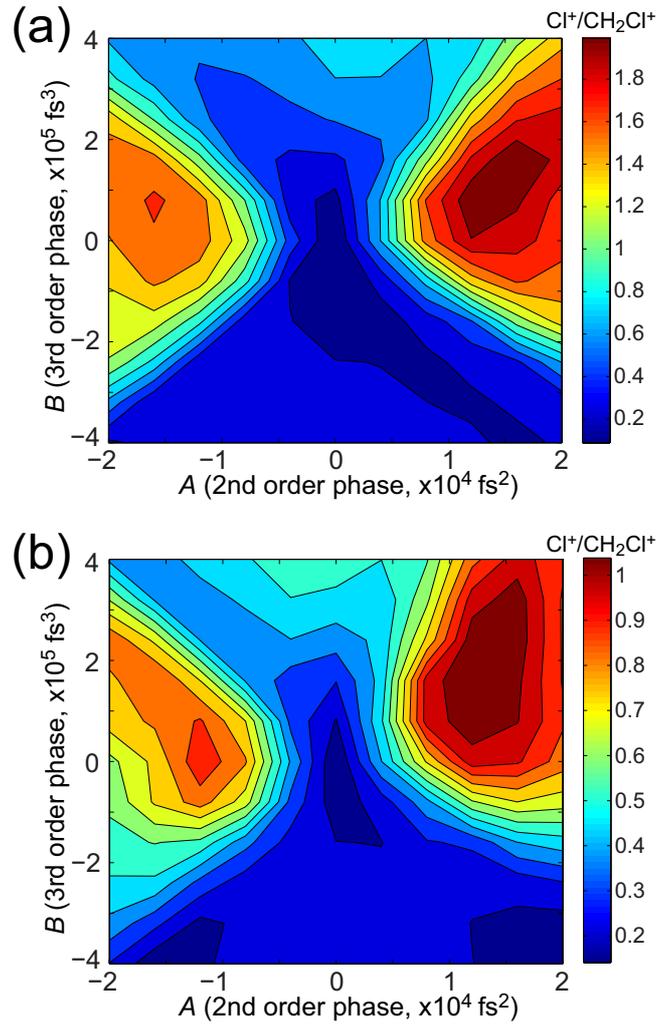}
\end{center}
\caption{\footnotesize Ratio Cl$^+$/CH$_2$Cl$^+$ from \brcl\ms as a function of second and third order spectral phase coefficients $A$ and $B$, which produces a two-dimensional {\it control landscape} \cite{mike1}. (a) Control landscape recorded on System I. (b) Control landscape recorded on System II. The absolute yield of the ratio for each landscape is denoted by the color bars to the right of the plots in (a) and (b). The general shape of these control landscapes is similar, with the two maxima occurring in nearly the same regions. The discrepancy in absolute yields is attributed to the lower laser intensity on System II.\label{landscape1}}
\end{figure}

The objective ratio Cl$^+$/CH$_2$Cl$^+$ (c.f., Table \ref{fit2}) from \brcl\ms as a function of coefficients $A$ and $B$ is plotted in Figure \ref{landscape1}(a) for System I and Figure \ref{landscape1}(b) for System II. The plots in Figure \ref{landscape1} represent {\it control landscapes} \cite{mike1}, or the functional relationship between an objective value (here the ratio Cl$^+$/CH$_2$Cl$^+$) with the restriction here to two variables chosen to construct the laser pulse. The two control landscapes have the same general features, with globally minimal objective yields around the origin (0,0) corresponding to the TL pulse, and two disconnected maxima, with the global maximum corresponding to positive values of $A$ and $B$ on both landscapes. The asymmetry in the recorded control landscapes with respect to negative and positive values $A$ and $B$ reflects the dependence of the control objective on the temporal profile of the shaped laser pulse, as was observed in Refs. \cite{melab1,melab2}.

While the two landscapes in Figure \ref{landscape1} are qualitatively similar, the maximum objective yield is nearly twice as high on System I. This discrepancy is attributed to the laser intensity being higher for System I, as discussed in Section \ref{int}. It was found that constructing the same control landscape through spectral phase variation while limiting the pulse energy reduced the attainable objective ratio Cl$^+$/CH$_2$Cl$^+$ in experiments conducted on System I (not shown), which suggests that increased laser intensity makes it possible to produce greater objective yields. Nevertheless, the qualitative similarity of the two landscapes shows that the molecular response to particular LCM masks is reasonably robust across the two laser systems. This result provides a basis to expect some degree of effective transfer when employing optimized photonic reagents.

\subsection{Efficacy of transferred optimal photonic reagents}\label{opt}
The experiments in this section assess the extent to which the optimal LCM masks identified on System I may be applied to the same compound and still be effective when implemented on System II. Given the results in Section \ref{param} that a higher absolute objective yield was obtained with the same photonic reagents on System I due to increased laser intensity, it is expected that absolute objective yields under interaction with the {\it optimal} photonic reagents may be lower on System II. Thus, we measure the {\it relative} efficacy of the System I photonic reagents applied to System II as compared to optimal photonic reagents identified on System II itself to assess the success of photonic reagent transfer.

\begin{figure}[htbp]
\begin{center}
\includegraphics[width=8.5cm]{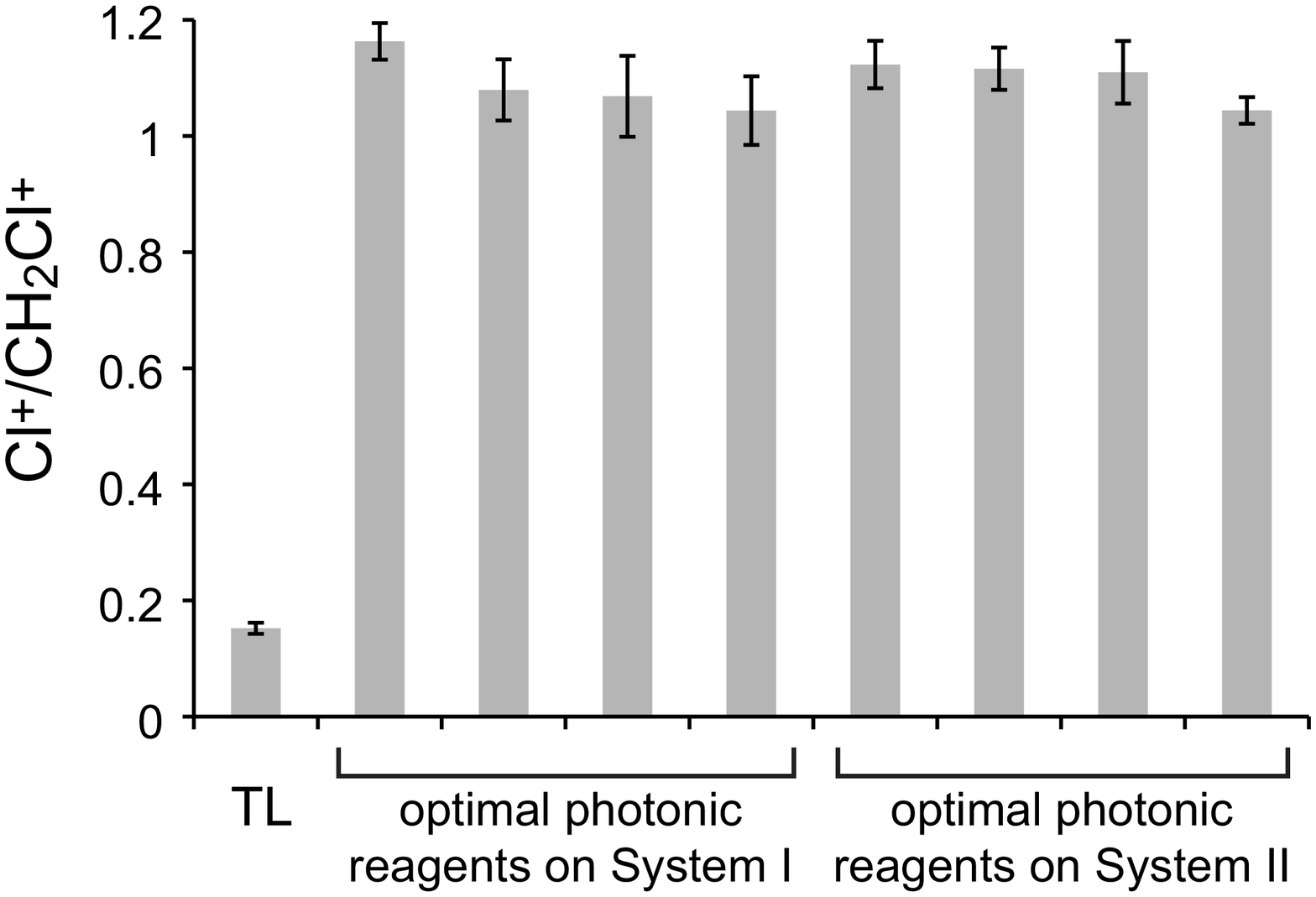}
\end{center}
\caption{\footnotesize Ratios Cl$^+$/CH$_2$Cl$^+$ from \brcl\ms obtained from four optimal photonic reagents identified with System I (left) and with System II (right), all measured on System II. For reference, the yield obtained with the TL pulse is shown. Error bars denote the standard deviation about the mean yield. The yields from the System I and System II photonic reagents are nearly the same, indicating that the photonic reagents optimal for System I are also optimal for System II. \label{brcl}}
\end{figure}

For \brcl, four photonic reagents producing an optimal ratio of Cl$^+$/CH$_2$Cl$^+$ were identified on System I \cite{melab2}. Subsequently, four photonic reagents that optimize the same objective ratio were identified on System II using the methods in Section \ref{alg}. The product ratios $^{35}$Cl$^+$/CH$_2\thinspace^{35}$Cl$^+$ produced from these eight photonic reagents were measured on System II and are shown in Figure \ref{brcl}(a), along with the ratio produced from the TL pulse, which is significantly lower ($^{35}$Cl$^+$/CH$_2\thinspace^{35}$Cl$^+\simeq0.15$) than for the optimized photonic reagents with $^{35}$Cl$^+$/CH$_2\thinspace^{35}$Cl$^+\simeq1.1$. The error bars denote the standard deviation about the mean yield measured from 100 samples of 1000 averaged laser shots. For reference, the optimal product ratio measured on System I for the four photonic reagents from Ref. \cite{melab2} was $^{35}$Cl$^+$/CH$_2\thinspace^{35}$Cl$^+\simeq 3.0$. Thus, the yields obtained on System II in Figure \ref{brcl} are significantly lower than on System I, as may be expected due to the lower laser intensity. Nevertheless, in the experiments conducted on System II, the yields obtained from the photonic reagents identified on System I are nearly indistinguishable from the yields of the photonic reagents identified on System II, indicating that the System I photonic reagents are also optimal when implemented on System II.

\begin{figure}[htbp]
\begin{center}
\includegraphics[width=8.5cm]{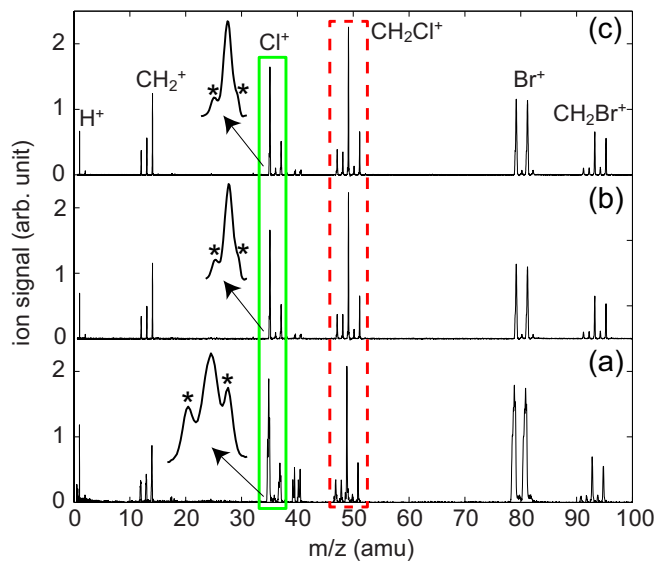}
\end{center}
\caption{\footnotesize TOF spectra of \brcl\ms obtained from (a) System I identified photonic reagent applied on System I, (b) System I identified photonic reagent applied on System II, and (c) System II identified photonic reagent applied on System II. The scale of the ion signal is the same for all three plots. The Cl$^+$ signals (in green solid-line box) are magnified as shown by the arrows. Peaks marked by an asterisk $*$ arise from Coulomb explosion. The red dashed-line box encloses the CH$_2$Cl$^+$ signal to be minimized. All spectra look similar, with the exception of significantly enhanced Coulomb explosion measured on System I in (a), which produces a higher integrated yield of Cl$^+$.\label{tofoldnew}}
\end{figure}

Comparison of the TOF spectra of \brcl\ms produced from the optimal photonic reagents identified on Systems I and II reveals the source of enhanced absolute objective yields obtained on System I. Figure \ref{tofoldnew} shows the spectra (a) measured on System I from a photonic reagent optimized on System I, (b) from the same System I photonic reagent but now measured on System II, and (c) from an optimal photonic reagent identified and measured on System II. The Cl$^+$ ion signal to be maximized is indicated by the green (solid line) boxes, and the CH$_2$Cl$^+$ signal to be minimized is indicated by the red (dashed line) boxes. The overall structure of the spectra are similar, with the relative height of the Cl$^+$ signal being only slightly lower than the corresponding CH$_2$Cl$^+$ signals for spectra measured on System II in Figure \ref{tofoldnew}(b) and (c) as compared to Figure \ref{tofoldnew}(a). However, magnification of the Cl$^+$ signal, shown to the left of the green box in each spectrum, shows a significant increase in the ion yield from Coulomb explosion (peaks marked with an asterisk) for the spectrum measured on System I in Figure \ref{tofoldnew}(a), which increases its total integrated yield. The enhanced absolute objective yield on System I is thus due to a significantly larger contribution of Coulomb explosion formation of Cl$^+$, as expected from the higher peak intensity on System I since Coulomb explosion is an intensity-dependent process \cite{gibson}. Nevertheless, the similarity between the spectra in Figure \ref{tofoldnew}(b) and (c) shows that the photonic reagents identified on both Systems I and II produce nearly identical objective yields when measured on System II (c.f., Figure \ref{brcl}), thus demonstrating effective transfer of the System I photonic reagents to System II.

The discrepancy between the center spectral wavelengths on Systems I and II raises the question of whether shifting the LCM masks identified on System I to line up with the pixel corresponding to 800nm on the System II LCM (from pixel 320 on the System I LCM to pixel 362 on the System II LCM) might improve the yields from the System I photonic reagents measured on System II. This procedure was found to produce at most a small $\lesssim10\%$ enhancement in objective yields for the four optimal photonic reagents identified on System I for \brcl\ms when measured on System II. The remaining experiments in this work are therefore conducted without the latter spectral shifting. However, other types of optimal control experiments, particularly involving resonances in the quantum system, may depend more strongly on the detailed spectral properties of the shaped laser pulse. Thus, shifting and rescaling the LCM masks to line up with the laser spectrum may be necessary for effective transfer in such cases. 

\begin{figure}[htbp]
\begin{center}
\includegraphics[width=8.5cm]{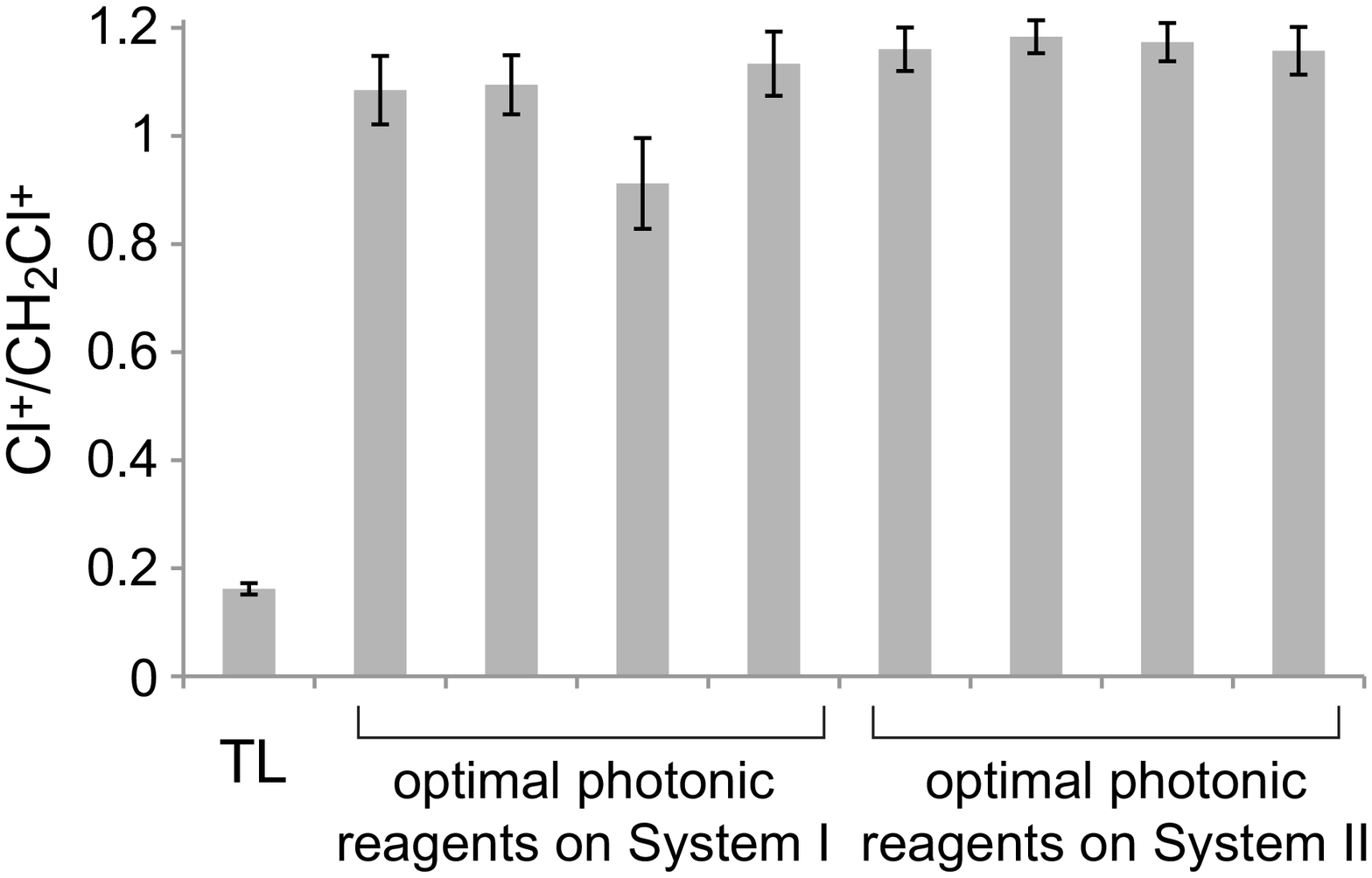}
\end{center}
\caption{\footnotesize Ratios $^{35}$Cl$^+$/$^{35}$CH$_2$Cl$^+$ from \icl\ms obtained from four optimal photonic reagents identified with System I (left) and with System II (right), all measured on System II. For reference, the yield obtained with the TL pulse is shown. Error bars denote the standard deviation about the mean yield. The yields from the System I and System II photonic reagents are nearly the same.\label{icl}}
\end{figure}

The good relative efficacy of the photonic reagents identified on Systems I and II for \brcl\ms opens up exploring whether optimal photonic reagents for other substrates are also transferrable. To investigate this proposition, we performed the same photonic reagent transfer experiment on \icl. Four System I photonic reagents, as well as four new photonic reagents discovered on System II, were applied to \icl\ms on System II to test their efficacy on maximizing the product ratio $^{35}$Cl$^+$/$^{35}$CH$_2$Cl$^+$. The results in Figure \ref{icl} show that three of the four System I photonic reagents successfully produce at least $\sim95\%$ of the optimal objective yield obtained from optimization of the photonic reagents on System II. In contrast, difficulties for transfer arose when considering an optimal photonic reagent for the objective I$^+$/CH$_2$I$^+$ from \ii\ms identified on System I. In Ref. \cite{melab2}, the optimization of this objective was unstable due to operating near the threshold of $S_2$ in Eq. (\ref{obj2}) for the CH$_2$I$^+$ signal. This situation shows that effective photonic reagent transfer can only be expected for reasonably robust objective yields.

\subsection{Trends across a homologous chemical family}\label{homo}
\begin{figure}[htbp]
\begin{center}
\includegraphics[width=8.5cm]{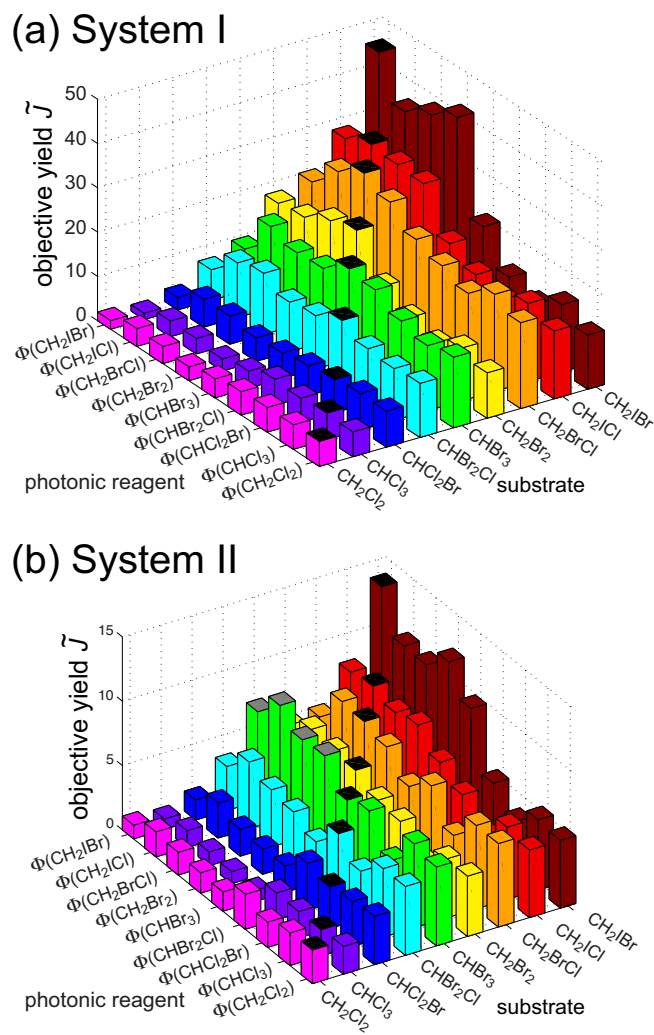}
\end{center}
\caption{\footnotesize Three-dimensional plots showing objective yield  from 81 distinct photonic reagent-substrate combinations as a function of both the photonic reagents and substrates: (a) photonic reagents identified and measured on System I; (b) the same set of System I identified photonic reagents applied to the substrates on System II. On both plots, the ``photonic reagent'' axis denotes the molecule on which the photonic reagent was optimized. The bars are colored by the compound on the ``substrate'' axis denoting the substrate for which the objective yield was measured: \cll\ms (magenta), \clll\ms (violet), \brcll\ms (blue), \brrcl\ms (cyan), \brrr\ms (green), \brr\ms (yellow), \brcl\ms (orange), \icl\ms (red), and \ibr\ms (maroon); the height of the colored bars indicates the objective yield $\tilde{J}$ in Eq. (\ref{normtl2}). The black tops of the bars on the diagonal denote the cases of direct optimization, i.e., photonic reagent = substrate, which generally produce the best objective yields for each substrate molecule among all photonic reagents. The exception for \brrr\ms in (b) is indicated by the grey tops on the bars and explained in the text. \label{allmolecules}}
\end{figure}

In Ref. \cite{melab2}, photonic reagents optimized on {\it each} halomethane compound were applied to the {\it remaining} compounds in the family in order to identify systematic trends in the objective yield as a function of chemical composition. To quantify the similarity between compounds in the molecular family, they were ordered in Figure \ref{allmolecules} by the sequence in Table \ref{fit2} according to the increasing product ratio of halogen ion/methyl halide ion (i.e., indicated by the bars with a black top on Figure \ref{allmolecules}). This sequence is in accord with natural electronic variability (e.g., polarizability) of the halogens on the parent molecules. As each optimal photonic reagent was tested with all compounds, the objective yields form a total of 81 photonic reagent-molecular substrate interactions. All objective yields $\tilde{J}$ reported in Figure \ref{allmolecules} are normalized to the yield with the 360$\mu$J TL pulse, i.e.,
\begin{equation}
\tilde{J}=\frac{{S_{1}}/{S_2}}{{S_{1}(\text{TL})}/{S_2(\text{TL})}},\label{normtl2}
\end{equation}
where $S_1(\text{TL})$ and $S_2(\text{TL})$ denote the ion signals measured with the 360$\mu$J TL pulse such that $\tilde{J}(\text{TL})=1.0$ (c.f., $A=B=C=0$ in Eq. (\ref{obj2})).

In Ref. \cite{melab2}, the objective yields $\tilde{J}$ were visualized by the three-dimensional plot shown in Figure \ref{allmolecules}(a). The ``photonic reagent" axis indicates the substrate on which the photonic reagent $\Phi(\text{substrate})$ was optimized. The substrate producing each recorded objective yield is denoted on the ``substrate" axis, with the color of the bars indicating the substrate. The objective yield in Eq. (\ref{normtl2}) from each photonic reagent and substrate combination is plotted as the height of the corresponding colored bar. The black tops of the bars on the diagonal in Figure \ref{allmolecules}(a) indicate objective yields obtained by direct optimization (i.e., photonic reagent-trained compound = substrate compound). Two trends observed in Figure \ref{allmolecules}(a) are of particular interest: (i) the objective yield increases according to the ordering of the compounds with the halogen composition varying from Cl $\to$ Br $\to$ I, and (ii) a systematic decrease in objective yields away from the diagonal is observed, e.g., the optimal photonic reagent for \brcl\ms is effective for its neighbor \icl, but not very effective for the more remote compound \clll.

The photonic reagents from System I used to construct Figure \ref{allmolecules}(a) were applied to System II to produce the analogous plot in Figure \ref{allmolecules}(b). The plots in Figures \ref{allmolecules}(a) and (b) are qualitatively similar despite the reduced normalized objective yields in Figure \ref{allmolecules}(b). Furthermore, the two trends (i) and (ii) above are observed in Figure \ref{allmolecules}(b) as well, which indicates qualitative successful transfer of the System I photonic reagents to System II. However, distinct behavior is observed in \brrr\ms because the photonic reagents optimized on \brcl, \icl, \brr, and \ibr\ms produce better objective yields for \brrr\ms in Figure \ref{allmolecules}(b) than from direct optimization, as indicated by the grey tops of the colored bars. This result arises from specifying a floor threshold signal for $S_2$ in Eq. (\ref{obj2}). On System I, the $S_2$ signal in \brrr\ms upon interaction with the latter photonic reagents was below the threshold, and the objective yields in Figure \ref{allmolecules}(a) were reported with $S_2$ being the threshold value. However, on System II, the $S_2$ signal was never below the threshold upon interaction with any photonic reagent, producing apparently higher objective yields from \brrr\ms with the latter photonic reagents in Figure \ref{allmolecules}(b). Since the photonic reagent generated for \brrr\ms was not robust on System I, this case illustrates the danger of attempting to perform transferability when operating with non-robust photonic reagents.

\section{Discussion and conclusion}\label{disc}
This work examined the extent to which photonic reagents may be transferred from one laser system to another and still produce optimal objective yields for halomethane fragmentation. A summary of the transferability tests performed for the present TOF-mass spectral experiments is given in Table \ref{sum}. Effective photonic reagent transfer was demonstrated by the qualitatively reproduced trends observed in Sections \ref{param} and \ref{homo}, as well as equivalent quantitative objective yields measured on System II in \brcl\ms and \icl\ms in Section \ref{opt}. { All three investigations showed that photonic reagents identified on System I could be effectively transferred to System II in the context of producing similar yields of the target control objectives. This practical success demonstrated that sufficient care was exercised in setting and comparing the operating parameters of both lasers, as presented in Section \ref{expt}.} Such laser characteristics and optics issues can be readily measured in any laboratory before attempting a transfer of control pulse shapes. { Although additional pulse characterization techniques could provide a more rigorous comparison of the produced photonic reagents, the results in the present work show that basic measurements of laser parameters are sufficient in the context of controlled molecular fragmentation.} 

The effective transfer of LCM masks between laser systems suggests that photonic reagents can operate in an analogous way to chemical reagents \cite{melab1,melab2}. The latter investigations in Refs. \cite{melab1,melab2} showed systematic trends in photoproduct yields of halomethanes as a function of both photonic reagent and substrate characteristics to enable a qualitative comparison of the behavior of photonic reagents and chemical reagents upon interaction with a set of chemically homologous substrates. The current results further support the analogy between photonic and chemical reagents by now showing the additional feature of successful transfer of photonic reagents to produce qualitatively similar product yields when implemented in a distinct experimental setup. The present results and those in Refs. \cite{melab1,melab2} establish a clear foundation for considering photonic reagent ``chemistry''. 

Effective photonic reagent transfer between laser systems is important for realizing practical applications of experimental quantum control. For example, applying optimal dynamic discrimination (ODD) of similar molecules \cite{roth} ideally should produce a reliable detection signal, regardless of the test location. Technical challenges for photonic reagent transfer beyond the ones discussed in this work may impact the degree to which transfer is effective for control objectives other than molecular fragmentation. In particular, consideration of the spectral output including the center wavelength, bandwidth, etc. may call for special care to obtain effective transfer of the photonic reagents. As the physical meaning of a photonic reagent is determined by only a modest number of parameters describing the laser field at the spot where light-matter interaction takes place, a standardized laser and pulse shaper would be desirable to effectively characterize photonic reagents. We hope that the present proof-of-principle demonstration also promotes further investigation into the ability of transferring photonic reagents for other optimal control objectives to bring the dreams of practical optimal control applications closer to reality. 

\renewcommand{\baselinestretch}{1}

\renewcommand{\baselinestretch}{1}
\begin{table}[htbp]
\begin{center}
\begin{tabular}{|c|c|c|}
\hline
Parameter&System I&System II\\\hline\hline
pulse duration&25-30fs&25-30fs\\\hline
pulse bandwidth&55-60nm&48-55nm\\\hline
pulse energy&2mJ&1.9mJ\\\hline
repetition rate&1kHz&3kHz\\\hline
$\lambda_0$&800nm&791nm\\\hline
\end{tabular}
\end{center}
\caption{\footnotesize List of laser parameters on Systems I and II. The values recorded denote their ranges reported under typical operating conditions of each laser system. No attempt was made to make any of the parameters match more closely. The pulse energies refer to the total amplifier output, only a portion of which was sent to the shaper. The pulse energies out of the shaper are comparable (c.f., Table \ref{param3}).\label{param1}}
\end{table}

\begin{table}[htbp]
\begin{center}
\begin{tabular}{|c|c|c|}
\hline
Parameter&System I&System II\\\hline\hline
center pixel&320&320\\\hline
grating resolution&1600 lines/mm&1400 lines/mm\\\hline
nm/pixel&0.155&0.179\\\hline
\end{tabular}
\end{center}
\caption{\footnotesize List of laser pulse shaper parameters for Systems I and II. \label{param2}}
\end{table}

\begin{table}[htbp]
\begin{center}
\begin{tabular}{|c|c|c|}
\hline
Parameter&System I&System II\\\hline\hline
path length from shaper&360cm&345cm\\\hline
pulse energy (before entering vacuum chamber)&350-400$\mu$J&330-380$\mu$J\\\hline
focal spot size&$40\mu$m&45$\mu$m\\\hline
estimated intensity&$930-1270$ TW/cm$^{2}$&$670-960$ TW/cm$^2$\\\hline
\end{tabular}
\end{center}
\caption{ \footnotesize List of experimental parameters for shaped pulse delivery on Systems I and II. The expected intensity ranges are estimated based on the lower and upper limits of pulse energy and duration from Tables \ref{param1} and \ref{param2} with the focal spot size. The intensity is expected to be $\sim 20\%$ higher on System I based on the parameter differences between the two laser systems. \label{param3}}
\end{table}

\begin{table}[htpb]
\begin{center}
\begin{tabular}{|c|c|}
\hline
species&optimized ratio\\\hline\hline
\cll&Cl$^+$/CH$_2$Cl$^+$\\\hline
\clll&Cl$^+$/CHCl$_2^+$\\\hline
\brcll&Cl$^+$/CHCl$_2^+$\\\hline
\brrcl&Cl$^+$/CHBrCl$^+$\\\hline
\brrr&Br$^+$/CHBr$_2^+$\\\hline
\brr&Br$^+$/CH$_2$Br$^+$\\\hline
\brcl&Cl$^+$/CH$_2$Cl$^+$\\\hline
\icl&Cl$^+$/CH$_2$Cl$^+$\\\hline
\ibr&Br$^+$/CH$_2$Br$^+$\\\hline
\end{tabular}
\end{center}
\caption{\footnotesize Control objectives for each compound. The ratio of the integrated ion signals from each mass fragment are used to evaluate the objective function $J$ in Eq. (\ref{obj2}), and the corresponding objective yields reported in Figure \ref{allmolecules}. \label{fit2}}
\end{table}

\begin{table}[htbp]
\begin{center}
\begin{tabular}{|c|c|c|}
\hline
Section&measurement&result\\\hline\hline
3.2&control landscape&similar shape, Figure \ref{landscape1}\\\hline
3.3&compare optimal&similar yield measured\\
&photonic reagents&on System II for photonic\\
&&reagents from Systems I and II,\\
&&Figures \ref{brcl}, \ref{tofoldnew}, and \ref{icl}\\\hline
3.4&compare systematic trends&similar trends with\\
&in objective yields&minor exceptions, Figure \ref{allmolecules}\\\hline
\end{tabular}
\end{center}
\caption{\label{sum} \footnotesize Summary of the successful measurements taken on Systems I and II in order to assess the transferability of photonic reagents. The objective yields measured on System II, however, were lower for {\it all} photonic reagents due to the lower laser intensity.}
\end{table}

\newpage
\renewcommand{\baselinestretch}{1}

\newpage

\newpage

\newpage

\newpage

\newpage

\newpage
\end{document}